\documentclass{aastex631}

\usepackage[utf8]{inputenc}
\usepackage{CJK}
\usepackage{amsmath}

\usepackage{booktabs}
\usepackage{tablefootnote}
\usepackage{multirow}

\newcommand{\ghz}{\ensuremath{\,{\rm GHz}}}
\newcommand{\mum}{\ensuremath{\,{\rm \mu m}}}

\newcommand{\K}{\ensuremath{\,{\rm K}}}

\newcommand{\s}{\ensuremath{\,{\rm s}}}
\newcommand{\g}{\ensuremath{\,{\rm g}}}

\newcommand{\cmg}{\ensuremath{\,{\rm cm^2\,g^{-1}}}}

\newcommand{\spitzer}{\emph{Spitzer}}

\newcommand{\herschel}{\emph{Herschel}}
\newcommand{\planck}{\emph{Planck}}

\shorttitle{FIR dust opacity in dark clouds}
\shortauthors{Li et al.}

\begin{document}
\begin{CJK*}{UTF8}{gbsn}

\title{Spatial Variations of Dust Opacity and Grain Growth in Dark Clouds: L1689, L1709 and L1712}

\author[0000-0001-9328-4302]{Jun Li (李军)}
\affiliation{Center for Astrophysics, Guangzhou University, Guangzhou 510006, People's Republic of China}
\correspondingauthor{Jun Li}
\email{lijun@gzhu.edu.cn}

\author[0000-0003-3168-2617]{Biwei Jiang (姜碧沩)}
\affiliation{Institute for Frontiers in Astronomy and Astrophysics, Beijing Normal University, Beijing 102206,  People's Republic of China}
\affiliation{Department of Astronomy, Beijing Normal University, Beijing 100875, People's Republic of China}

\author[0000-0003-2645-6869]{He Zhao (赵赫)}
\affiliation{Purple Mountain Observatory and Key Laboratory of Radio Astronomy, Chinese Academy of Sciences, 10 Yuanhua Road, Nanjing 210033, People's Republic of China}

\author[0000-0002-5435-925X]{Xi Chen (陈曦)}
\affiliation{Center for Astrophysics, Guangzhou University, Guangzhou 510006, People's Republic of China}
\correspondingauthor{Xi Chen}
\email{chenxi@gzhu.edu.cn}

\author[0000-0002-2895-7707]{Yang Yang}
\affiliation{Center for Astrophysics, Guangzhou University, Guangzhou 510006, People's Republic of China}



\begin{abstract}
	
The far-infrared (FIR) opacity of dust in dark clouds within the Ophiuchus molecular cloud is investigated through multi-wavelength infrared observations from UKIDSS, \spitzer\ and \herschel. Employing the infrared color excess technique with both near-infrared (NIR) and mid-infrared (MIR) photometric data, a high-resolution extinction map in the $K$ band ($A_K$) is constructed for three dark clouds: L1689, L1709, and L1712. The derived extinction map has a resolution of $1'$ and  reaches a depth of $A_K\sim3$ mag. The FIR optical depths $\tau_{250}$ at a reference wavelength of $250\mum$ are obtained by fitting the \herschel\ PACS and SPIRE continuum data at 100, 160, 250, 350 and 500 \mum\ using a modified blackbody model. The average dust opacity per unit gas mass at $250\mum$, $r\kappa_{250}$ is determined through a pixel-by-pixel correlation of $\tau_{250}$ with $A_K$, yielding a value of approximately $0.09\cmg$, which is about 2-3 times higher than the typical value in the diffuse interstellar medium (ISM). Additionally, an independent analysis across 16 sub-regions within the Ophiuchus cloud indicates spatial variations in dust opacity, with values ranging from 0.07-0.12\cmg. Although the observed trend of increasing dust opacity with higher extinction implies grain growth, our findings indicate that rapid grain growth clearly not yet occurred in the dark clouds studied in this work.

\end{abstract}

\keywords{Dark interstellar clouds(352) --- Interstellar dust(836) --- Dust continuum emission(412) --- Interstellar extinction(841)}


\section{Introduction} \label{sec:intro}

Dark clouds within the Milky Way represent a subset of molecular clouds that remain visually opaque due to their high gas and dust density \citep{lynds1963}. These clouds, particularly their cores, are often the birthplaces of new stars and accompanying planetary systems. Precise knowledge of the initial conditions, including density and temperature, within these dark clouds can considerably enhance our understanding of star formation processes. However, molecular hydrogen, a major component in star formation, remains undetectable under conditions of low temperature and high density. Additionally, molecules such as CO are often depleted within the cores of dark clouds, and the associated molecular emission (i.e. $^{12}$CO) is usually optically thick. As a result, the near-infrared (NIR) extinction of dust and its far-infrared (FIR) emission serve as important indicators of physical conditions within the dark cloud cores prior to its collapse into a star \citep{Lada2007}. Moreover, the study of dust emission properties is of significant value in obtaining more reliable mass estimates for molecular clouds and further refining the initial mass function (IMF) of star formation. Consequently, the investigation of dust properties within dark clouds is fundamental to understanding the processes governing star formation.

A fundamental property of dust is their ability to absorb short-wavelength photons and re-emit energy at longer wavelengths \citep{li2005}. This absorption and emission behavior is commonly characterized by the `opacity' parameter $\kappa_\nu$, expressed in units of \cmg. This parameter plays a crucial role in estimating dust mass by fitting the FIR and millimeter continuum emission \citep{Fanciullo2020,li2022}. However, achieving accurate measurements has been constantly challenging. Historical measurements of the opacity have shown discrepancies spanning more than two orders of magnitude \citep{clark2019,Bianchi2022}, and spatial variations have been observed across different environments \citep{roy2013,martin2012}. Moreover, the widely adopted models for dust evolution studies, such as those proposed by \cite{ossenkopf1994} and \cite{ormel2011}, present absorption coefficients that differ by over threefold. The assumptions underpinning these models may not be universally applicable across complex interstellar environments, ranging from cold, starless cloud cores to heated region in star clusters. Recent studies have suggested that the opacity provided by dust evolution models does not adequately account for the FIR emission of cloud cores \citep{webb2017}. As a result, it becomes essential to constrain the opacity through observational data.

The opacity can be investigated by considering the ratio of extinction in the optical/NIR bands to FIR emission \citep{bianchi2003,kramer2003,shirley2011,suutarinen2013}. On the one hand, the \herschel\ Space Observatory \citep{pilbratt2010} provides photometric observations of dust emission spanning wavelengths between 70-500\mum. This enhances our ability to probe the physical properties of dust at relatively high angular resolution and across great depths in molecular clouds, although the results are somewhat mitigated by degeneracies in dust optical depth and temperature. On the other hand, extinction is independent of temperature and serves as an effective proxy for tracing the column density of dust. However, the construction of an extinction map relies heavily on the statistical analysis of the number density of background stars. As a result, in dense dark cloud cores, the measurement of extinction maps requires high sensitivity and photometric depth \citep{webb2017}. Previous extinction maps derived from the 2MASS survey typically have a spatial resolution of around 3$'$ \citep{Lombardi2008,juvela2016}, which is insufficient for resolving the internal structure of dark cloud cores. Although many studies have reported an increase in opacity with increasing extinction, opacity in dense cloud cores tends to be underestimated due to the limited depth of $A_V<10$\,mag \citep{martin2012,roy2013,juvela2015,lewis2022}. To better study the dust properties of clouds and cores, it is essential to construct an extinction map with a spatial resolution on par with \herschel\ observations (i.e. $<1'$). The deeper and more sensitive NIR surveys from Visible and Infrared Survey Telescope for Astronomy (VISTA) and the United Kingdom Infrared Telescope (UKIRT) can help to derive such high-resolution extinction maps \citep{chu2021,Zhang2022}. 

In this work, we aim to investigate dust opacity by correlating NIR extinction with FIR emission within the Ophiuchus cloud, focusing specifically on the three dark clouds L1689, L1709, and L1712. These clouds are located at a distance of $\sim$139 pc \citep{Ortiz2018} with relatively clean background, making it a suitable target for studying dust evolution in molecular cloud. Previous works have shown evidence of dust growth in Ophiuchus cloud based on the infrared extinction law \citep{Chapman2009,li2023}. Here, we focus on the FIR dust opacity by combining NIR extinction and FIR emission.

This paper is organized as follows. Section \ref{sec:obser} describes the data briefly. Section \ref{sec:ana}  presents the data analysis approach to derive extinction map and FIR optical depths. Section \ref{sec:result} provides the determination of dust opacity and discussions. Section \ref{sec:summary} is the summary.

\section{Data} \label{sec:obser}

\subsection{UKIDSS Data} \label{subsec:ukidss}

The primary objective of this study is an investigation of dark clouds L1689, L1709 and L1712 within the Ophiuchus cloud. We use photometry in the $J$, $H$, and $K$ bands, obtained from the Galactic Clusters Survey (GCS) of the UKIRT Infrared Deep Sky Survey (UKIDSS) Release 11 \citep{lucas2008}. The observed region encompasses an area of approximately 2.6\textdegree$\times$2.3\textdegree. Detailed descriptions of the photometric system and its calibration are available in \cite{Hewett2006} and \cite{Hodgkin2009}, respectively.

The UKIDSS observations provide superior resolution and sensitivity in comparison to 2MASS survey, achieving limiting magnitudes that are about 3.5 mag deeper across each band. The UKIDSS data can be retrieved from the WFCAM Science Archive\footnote{http://surveys.roe.ac.uk/wsa/index.html}. To ensure high data quality and keep only stellar sources, the photometric errors for all $JHK$ bands are required to be less than 0.2 mag and values of `PSTAR' flag larger than 0.9. This results in limiting magnitudes of 20.9, 19.4, and 19.0 mag for the $J$, $H$, and $K$ bands, respectively.

\subsection{Spitzer Data} \label{subsec:spitzer}

We collected observations on Ophiuchus molecular cloud from the \spitzer\ Space Telescope, which is a part of the \spitzer\ Legacy Project ``From Molecular Cores to Planet-Forming Disks" (c2d) \citep{https://doi.org/10.26131/irsa421}. We use data across four mid-infrared (MIR) IRAC bands, namely [3.6], [4.5], [5.8], and [8.0]. Detailed information about data processing can be found in \cite{evans2003} and \cite{padgett2008}. The \spitzer\ c2d observations of Ophiuchus cloud cover an area of about 8 square degrees. Access to the \spitzer\ c2d data is facilitated through the archive on the \spitzer\ Science Center's website\footnote{https://irsa.ipac.caltech.edu/data/SPITZER/C2D/}. To obtain a high-quality data sample, the signal-to-noise ratio (SNR) of photometry is required to be larger than 5, and the `QType' flag to be exclusively `star'. Consequently, the photometric limiting magnitudes for the four IRAC bands were 17.9, 17.1, 16.3, and 15.2 mag, respectively.

\subsection{ Herschel Data} \label{subsec:herschel}

The FIR continuum emission data are obtained from the \herschel\ Space Observatory \citep{pilbratt2010} by its two primary instruments: the Photodetector Array Camera and Spectrometer (PACS) \citep{poglitsch2010} and the Spectral and Photometric Imaging Receiver (SPIRE) \citep{griffin2010}. The PACS data offer resolutions of 6$''$ at both 70 and 100 $\mu$m, and an approximate resolution of 12$''$ at 160 $\mu$m. In contrast, the SPIRE data provide spatial resolutions of 18.1$''$, 25.2$''$, and 36.9$''$ at 250, 350, and 500\mum, respectively. It is crucial to acknowledge that these resolutions can vary depending on the specific observation mode and the conditions under which the observations were made. The datasets are publicly accessible via the \herschel\ Science Archive\footnote{http://archives.esac.esa.int/hsa/whsa/}.


\section{Analysis} \label{sec:ana}

\subsection{NIR Extinction mapping} \label{subsec:ext}

The Ophiuchus cloud is populated with various translucent clouds, dense cores, and young stellar objects (YSOs).To probe the central dense regions of these clouds, we incorporated both UKIDSS NIR and \spitzer\ MIR photometric data. Following the methodology outlined by \cite{roman2009}, we employed a combined approach utilizing the Near-Infrared Color Excess (NICE) method proposed by \cite{lada1994} and the NICER method revised by \cite{lombardi2001} to estimate extinction. Both techniques necessitate the prior knowledge of intrinsic colors. The NICE method requires only one observed color (i.e., the difference between photometric magnitude across two wavelength bands), while the NICER method needs two observed colors. Generally, we use the following formula to calculate the extinction in the $K$ band (i.e., $A_K$):
\begin{equation}
    A_K = C_{\rm el}\times E(m_{\lambda_1}-m_{\lambda_2})=C_{\rm el}\times[(m_{\lambda_1}-m_{\lambda_2})_{\rm obs}-(m_{\lambda_1}-m_{\lambda_2})_0]
\end{equation}
where $E(m_{\lambda_1}-m_{\lambda_2})$ is the color excess in the two bands of $\lambda_1$ and $\lambda_2$, which is the difference between the observed color $(m_{\lambda_1}-m_{\lambda_2})_{\rm obs}$  and the intrinsic color $(m_{\lambda_1}-m_{\lambda_2})_0$. The term $C_{\rm el}$ is the coefficient to convert color excess  $E(m_{\lambda_1}-m_{\lambda_2})$ to extinction $A_K$, which depends on the assumed extinction law. We adopt the results of $C_{\rm el}$ from \cite{li2023} who studied the extinction law within the Ophiuchus cloud using the same data (i.e., UKIDSS and \spitzer\ data) as that of this work. The corresponding coefficients $C_{\rm el}$ are listed in Table \ref{tab:int_cor}.

The two photometry catalogs from UKIDSS and \spitzer\ are merged within a tolerance radius of 1 arcsec using the TOPCAT software. The UKIDSS-\spitzer\ common area catalog is analyzed separately and merged partially: only \spitzer\ sources are used when the UKIDSS\ sources are either unavailable or unreliable. The deatiled procedures are as follows: (1) if all three $JHK$ photometry of a source are available, the extinction is estimated using the NICER method; (2) if a source lacks $J$ photometry but has $HK$ photometry, the extinction is evaluated through $E(H-K)$ using the NICE method; (3) if a source lacks both $J$ and $H$ photometry, the extinction is estimated by $E(K-[3.6])$. This process is continued until only $E([4.5]-[5.8])$ is used.


\begin{table*}
\centering
\caption{Parameters Employed in the Estimation of Extinction Using the NICE and NICER Methods.}
\label{tab:int_cor}
\begin{tabular}{c c c c c c }
\hline \hline
Color &  $J-H$ & $H-K$ &$K-[3.6]$ & $[3.6]-[4.5]$ & $[4.5]-[5.8]$  \\
\hline
 Intrinsic Color$^a$ &  0.497$\pm$0.102 & 0.153$\pm$0.086 & 0.155$\pm$0.117 & 0.019$\pm$0.042 & 0.023$\pm$0.013  \\

$C_{\rm el}$$^b$ & 0.77 & 1.21 &2.37 &12.05 &10.31  \\

 
\hline
\end{tabular}
\tablecomments{(a) Intrinsic colors from TRILEGAL simulations. Errors are the standard deviation of the colors. \\
(b) The coefficients for converting color excess $E(m_{\lambda_1}-m_{\lambda_2})$ to extinction $A_K$ by adopting the extinction law from \cite{li2023}. }
\end{table*}

\subsubsection{Intrinsic Colors} \label{subsec:int_color}

Normally, intrinsic colors are approximated by averaging the observed color of background stars from a nearby low-extinction region (also known as a control field). Nonetheless, selecting a low-extinction region from \spitzer\ observation is complicated because it covers the main structure of the Ophiuchus cloud. To address this, we choose to leverage the TRILEGAL model in simulating the stellar distribution in the Milky Way\footnote{http://stev.oapd.inaf.it/cgi-bin/trilegal} \citep{girardi2005}. In the TRILEGAL simulations, we designated a field of view with 2 square degrees directed towards the Ophiuchus cloud, set the extinction to zero, and keep all other parameters (i.e. the initial mass function, binary fraction, and Galaxy component) at their default values. We then selected the desired photometric system and set the $H$ band limiting magnitude to be 20 mag. Consequently, a stellar catalog was generated containing details such as stellar mass, temperature, and magnitudes in the corresponding bands. We then computed the average and $1\sigma$ dispersion of model colors as the estimates of the intrinsic colors and their associated uncertainties, respectively. For comparative purposes, we selected a low-extinction region from the UKIDSS observations, and estimated the average colors $(J-H)_0\approx0.495\pm0.293$\,mag and $(H-K)_0\approx0.163\pm0.266$\,mag. These values closely match the estimates from TRILEGAL simulations within $1\sigma$ uncertianty, as revealed in Table \ref{tab:int_cor}. This indicates that the TRILEGAL simulation can yield a reliable approximation of intrinsic colors. 

\begin{figure*}
	\centering
	\includegraphics[scale=0.45]{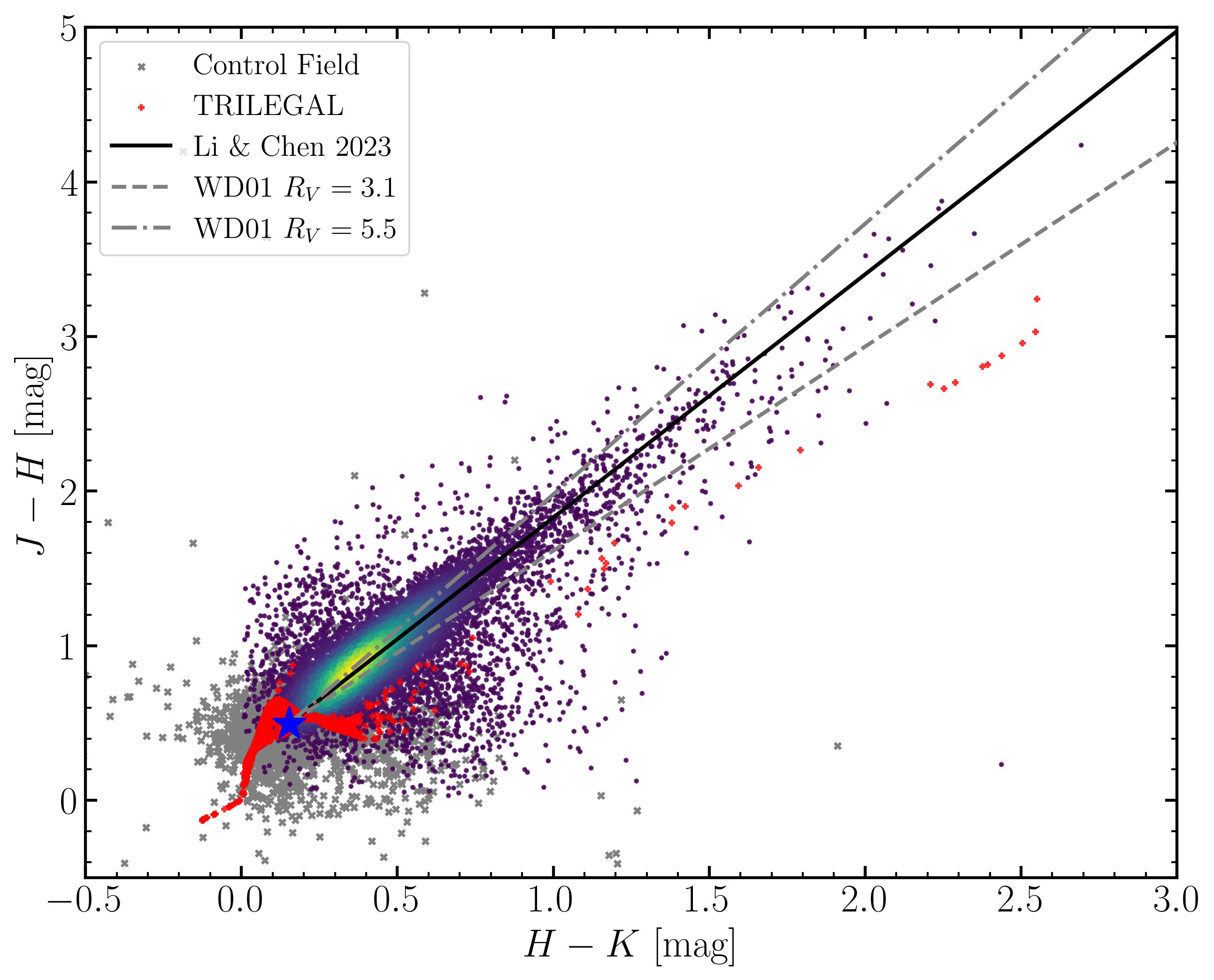}
	\caption{Color-color density distribution ($J-H$ vs. $H-K$) for stars within the regions L1689, L1709, and L1712. Stars from the control (zero-extinction) region and the TRILEGAL simulation are shown by grey and red markers, respectively. The blue star symbol represents the adopted intrinsic color $(J-H)_0$ and $(H-K)_0$. The black solid line presents the reddening law as measured by \cite{li2023}. For comparison, the reddening laws predicted by WD01 $R_V=3.1$ and $R_V=5.5$ models \citep{weingartner2001} are presented by the grey dashed and dot-dashed lines, respectively. 
		\label{fig:ccd_JH_Hk} }
\end{figure*}

Figure \ref{fig:ccd_JH_Hk} presents the color-color diagram (CCD) of observed $J-H$ versus $H-K$ for three dark clouds: L1689, L1709, and L1712. This figure also exhibits the distribution of stars as derived from the TRILEGAL simulations, as well as stars from the `low-extinction' field of the UKIDSS dataset. A noticeable shift of the stars from the dark clouds along the reddening vector in the CCD provides a strong indication of significant extinction present within these clouds.

\subsubsection{Map Construction}

Upon calculating the color excess and extinction $A_K$ for each individual star, we employed $A_K$ to generate an extinction map. The final extinction map is produced by utilizing a Gaussian filter with a Full Width at Half Maximum (FWHM) of 1 arcmin and a pixel size of 0.5 arcmin. The extinction value at each pixel is determined by the Gaussian weighted mean of extinction values for sources located within a radius three times the standard deviation ($3\sigma$) of the Gaussian beam. Thereafter, we calculated the errors associated with the mean extinction, represented as $\sigma_{A_K}$, using the conventional formula for a weighted average.

\begin{figure*}
	\centering
	\includegraphics[scale=0.53]{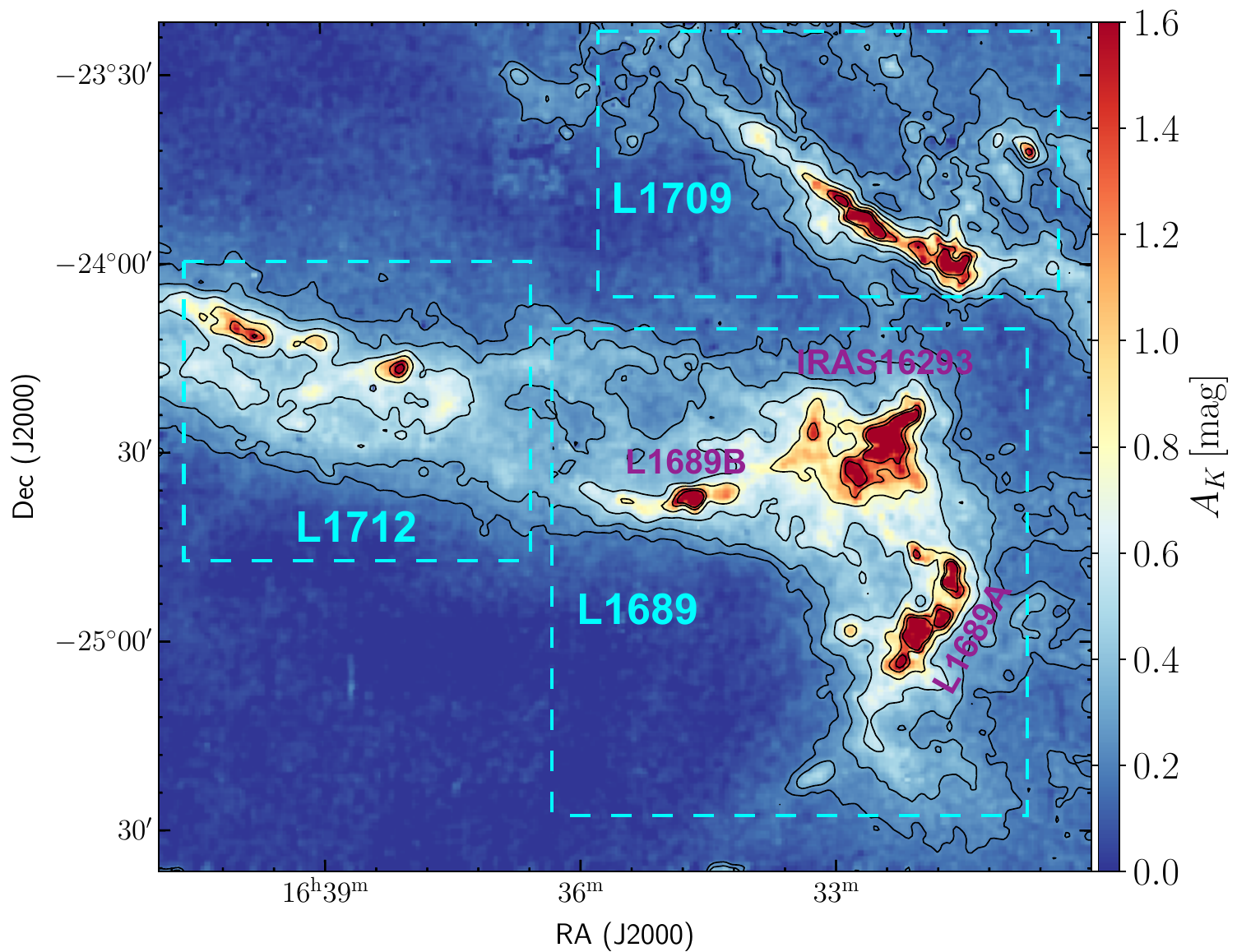}
	\caption{Extinction $A_K$ map of the eastern part of Ophiuchus cloud at a spatial resolution of $1'$. Contours are plotted at extinction levels of 0.2, 0.3, 0.5, 0.8, 1.2 and 1.5 mag. The cyan dashed rectangles indicate the locations of dark clouds: L1689, L1709 and L1712. Note that the color scale represents $A_K$ values only up to 1.6\,mag, dense regions with extinctions greater than 1.6\,mag remain unresolved in the current map.
		\label{fig:oph_ak_map} }
\end{figure*}

\begin{figure*}
	\centering
	\includegraphics[scale=0.5]{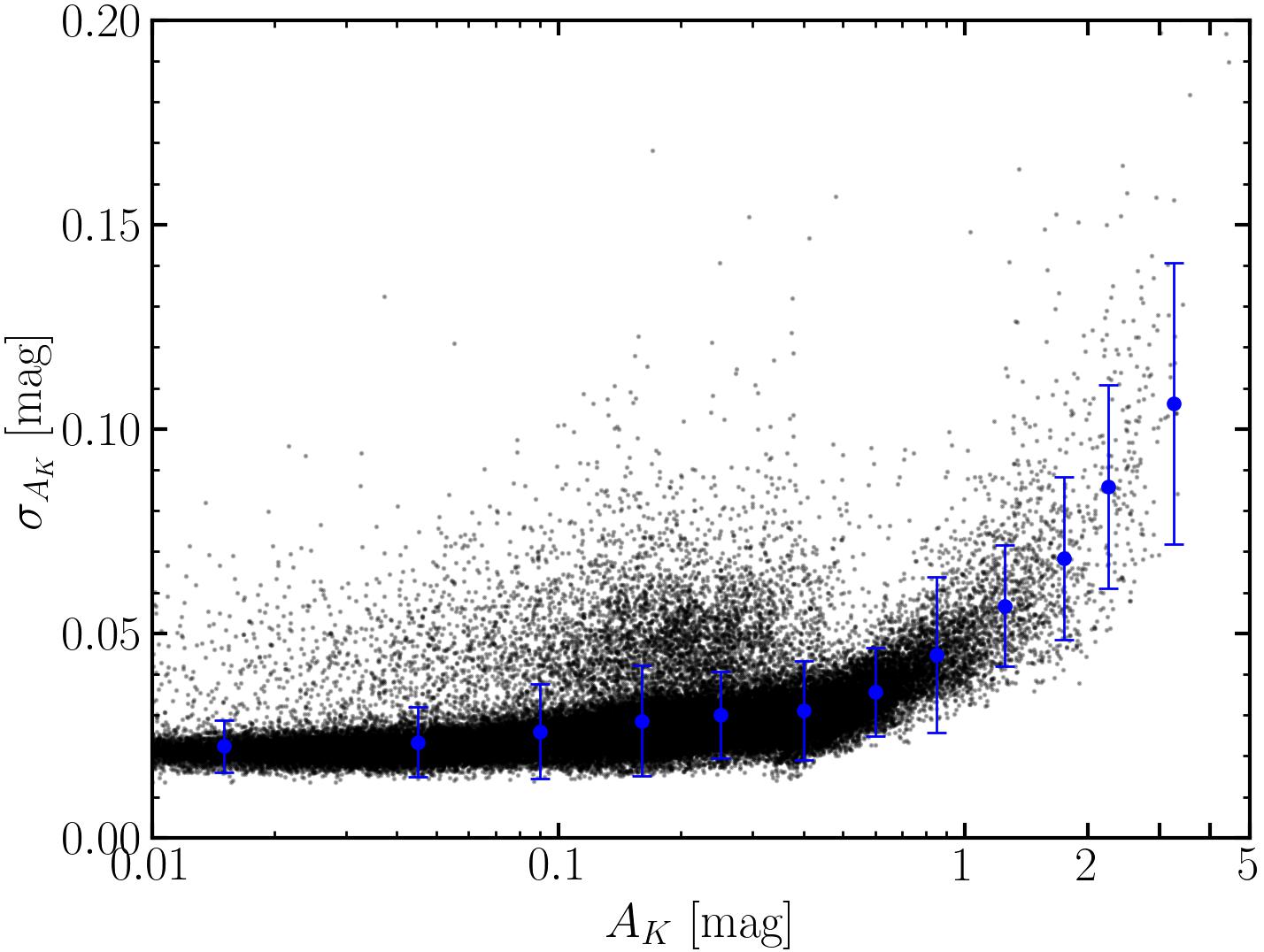}
	\caption{$\sigma_{A_K}$ vs. $A_K$ distribution for the extinction map of Figure \ref{fig:oph_ak_map}. The individual mean extinctions per pixel are shown in black dots. The blue dots with errorbars denote the median values and standard deviations, respectively.
	\label{fig:hist_ext} }
\end{figure*}

Figure \ref{fig:oph_ak_map} shows the resultant extinction map for the dark clouds L1689, L1709, and L1712  with a color scale showing $A_K$ up to 1.6 map. Figure \ref{fig:hist_ext} illustrates the relationship between the extinction error ($\sigma_{A_K}$) and $A_K$. For $A_K<0.5$ mag, $\sigma_{A_K}$ is relatively constant at approximately 0.02 mag, but it begins to increase for $A_K>0.5$ mag, reaching up to roughly 0.2 mag. Although the depth of extinction map reaches $A_K\approx3$ mag, it should be noted that regions with extinction $A_K>2$ mag are likely not resolved in Figure \ref{fig:oph_ak_map}. Previous measurements of the Ophiuchus extinction map derived from 2MASS data typically demonstrated a resolution of approximately 3 arcmin \citep[e.g.,][]{lombardi2001,juvela2016}. The extinction map presented in this study improve the resolution by a factor of 3, thereby enabling the detection of core structures within dark clouds and filaments.

\subsection{FIR SED Fitting\label{subsec:sed}}

\subsubsection{Zero-point Correction}

Due to the unknown thermal background of the instruments on Herschel, the flux measurements obtained from \herschel\  represent relative fluxes rather than absolute fluxes. In this work, we used the Level 2.5 data products for the \herschel\ PACS at 100\mum\ and 160\mum, as well as Level 3 data products for the \herschel\  SPIRE at 250\mum, 350\mum\ and 500\mum. The SPIRE Level 3 data were already zero-point corrected, however the PACS level 2.5 data have not undergone zero-point corrections. Therefore, before fitting the Spectral Energy Distribution (SED), it is imperative to correct the zero points of \herschel\ PACS data. Following the methodology commonly employed in prior studies \citep[e.g.][]{Lombardi2014,Singh2022}, we use the \planck\ data products derived from the \planck\ 2013 all-sky model of thermal dust emission \citep{planck2014}\footnote{The 2013 all-sky \planck\ data products were taken from https://wiki.cosmos.esa.int/planckpla/index.php/CMB\_and\_astro-physical\_component\_maps.}. These data provide three maps of dust temperature, ($T_{\rm d}$), dust emissivity index ($\beta$) and optical depth at 353\ghz\ ($\tau_{353}$), all of which were produced from an all-sky SED fit by a modified black-body model. The resolution of the $T_{\rm d}$ and $\tau_{353}$ maps is 5 arcmin, while the resolution of $\beta$ map is 30 arcmin. We then compute the fluxes at \herschel\ bands by convolving the \planck\ model SED with the \herschel\ filters at a resolution of 5 arcmin, denoted as $S_\nu^{\rm Planck}$. Subsequently, we perform a linear fit to the $S_\nu^{\rm Planck}$ and the observed \herschel\ fluxes convolved to the 5 arcmin resolution, $S_\nu^{\rm Herschel}$:
\begin{equation}
    S_\nu^{\rm Herschel} = a_\nu + b_\nu S_\nu^{\rm Planck}.
\end{equation}
The intercept $a_\nu$ from the linear fit offers an estimate for the absolute photometric calibration of \herschel\ observations, which represent the zero point of \herschel\ flux. Ideally, the slope $b_\nu$ of the linear fit should be close to 1. We estimate the errors of $a_\nu$ using a bootstrap method and present the results for the observations of L1689, L1709, and L1712 in Table \ref{tab:sed_cal}. There are substantial zero-point offsets in the PACS 100\mum\ and 160\mum\ images, ranging from several tens to hundreds of MJy/sr. This method has also been applied to the SPIRE data that have been previously corrected for zero-point offsets. The derived zero-point values for the SPIRE data were found to be near zero (less than 6 MJy/sr), demonstrating the effectiveness of this method in zero-point calibration.

\begin{table*}
\centering
\caption{Adopted Flux Calibrations, Color Corrections, zero points ($a_{\nu}$) and RMS noise at $1'$ resolution used for SED fitting.}
\label{tab:sed_cal}
\begin{tabular}{c c c  c c c c}
\hline \hline
Source & Calibration & 100\mum\ & 160\mum\ & 250\mum\ & 350\mum\ & 500\mum\\
\hline
 & Flux calibration  & 7\%   &  7\%  & 7\%   &  7\%   &    7\% \\
 & Color correction   & 1.09$\pm$0.10  &   1.01$\pm$0.05  &  1.02$\pm$0.02  &   1.01$\pm$0.01  &  1.03$\pm$0.02  \\
 \hline
L1709 & $a_\nu$ (MJy/sr) &  -71.17$\pm$0.90   &  -114.76$\pm$1.12  & 3.60$\pm$0.83 &  3.61$\pm$0.43  & 1.11$\pm$0.17 \\
 & $\rm RMS_{1'}$ (MJy/sr)     & 10.92   &  5.44  &  1.62  &  0.86 &   0.91\\
 \hline
 L1689 & $a_\nu$ (MJy/sr)  &  -121.05$\pm$0.84 &  -175.09$\pm$1.33 &   -3.43$\pm$0.79   &  0.09$\pm$0.43  &  0.37$\pm$0.26 \\
 & $\rm RMS_{1'}$ (MJy/sr)    & 10.71   &   5.26 &   1.65  & 0.88   &  1.11 \\
 \hline	
L1712 & $a_\nu$ (MJy/sr) & -102.07$\pm$0.75   &  -149.36$\pm$0.75 &  -5.20$\pm$0.17  &  -0.28$\pm$0.09  &  -0.12$\pm$0.22 \\
  & $\rm RMS_{1'}$ (MJy/sr)   & 10.88    &  5.35  &   1.65  &   0.88  &  1.11 \\
\hline
\end{tabular}
\end{table*}

\subsubsection{Maps of Optical Depth and Dust Temperature}

The zero-point corrected \herschel\ images were convolved to a resolution of 1 arcmin and regridded to a pixel size of 0.5 arcmin, consistent with the $A_K$ extinction map obtained in Section \ref{subsec:ext}. Given that dust thermal emission at the FIR and submillimeter wavelengths is typically optically thin, the surface brightness $I_\nu$ (in units of MJy/sr) of dust emission can be described by:
\begin{equation} \label{equ:sed}
I_\nu=B_\nu(T_{\rm d})(1-e^{-\tau_\nu})\approx \tau_\nu B_\nu(T_{\rm d})
\end{equation}
where $B_\nu(T_{\rm d})$ is the \planck\ function for dust temperature $T_{\rm d}$ and frequency $\nu$. $\tau_\nu$ is the dust optical depth, which can be expressed as:
\begin{equation}\label{equ:tau_kappa}
\tau_\nu = \mu m_{\rm H}N_{\rm H}r\kappa_\nu
\end{equation}
where $N_{\rm H}$ is the total hydrogen column density $N_{\rm H}=2N({\rm H_2})+N({\rm H})$, $\mu$ is the mean molecular weight (with $\mu=1.4$, assuming 10\% He),  and $r$ is the dust-to-hydrogen mass ratio $M_{\rm d}/M_{\rm H}$, which is presumed to be 1/100 and believed to vary minimally from diffuse to dense environment. $\kappa_\nu$ is the dust opacity, frequently refered to as the mass absorption coefficient or sometimes the emissivity. Another commonly characterized quantity is 
$\sigma_\nu \equiv \tau_\nu/N_{\rm H} = \mu m_{\rm H} r\kappa_\nu$, also called the opacity (the emission cross-section per H nucleon). In subsequent discussions, dust opacity is expressed as $r\kappa_\nu$, the opacity per unit gas mass. This can be converted to opacity per unit dust mass by simply dividing by the dust-to-hydrogen mass ratio $r$.

\begin{figure*}
	\centering
	\includegraphics[scale=0.55]{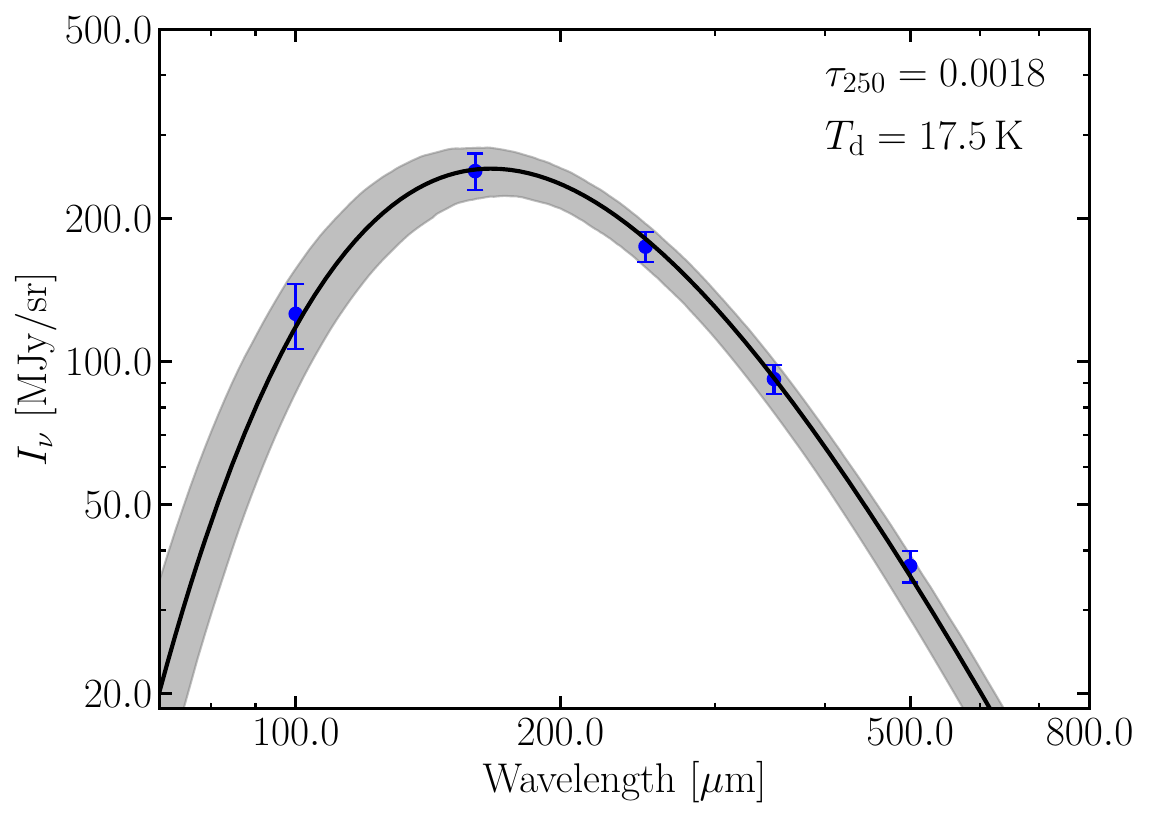}
	\caption{A SED example for a single spatial pixel within the Ophiuchus cloud using \herschel\ observations. The black solid line indicates the best-fit model with $\beta=2$ as described in Section \ref{subsec:sed}. The derived optical depth at 250\mum\ is $\tau_{250}=0.0018$ and dust temperature is $T_{\rm d}=17.5$\,K. The grey shaded area illustrates the 90\% confidence interval in MC analysis.
		\label{fig:sed_fit_example} }
\end{figure*}

\begin{figure*}
	\centering
	\includegraphics[scale=0.56]{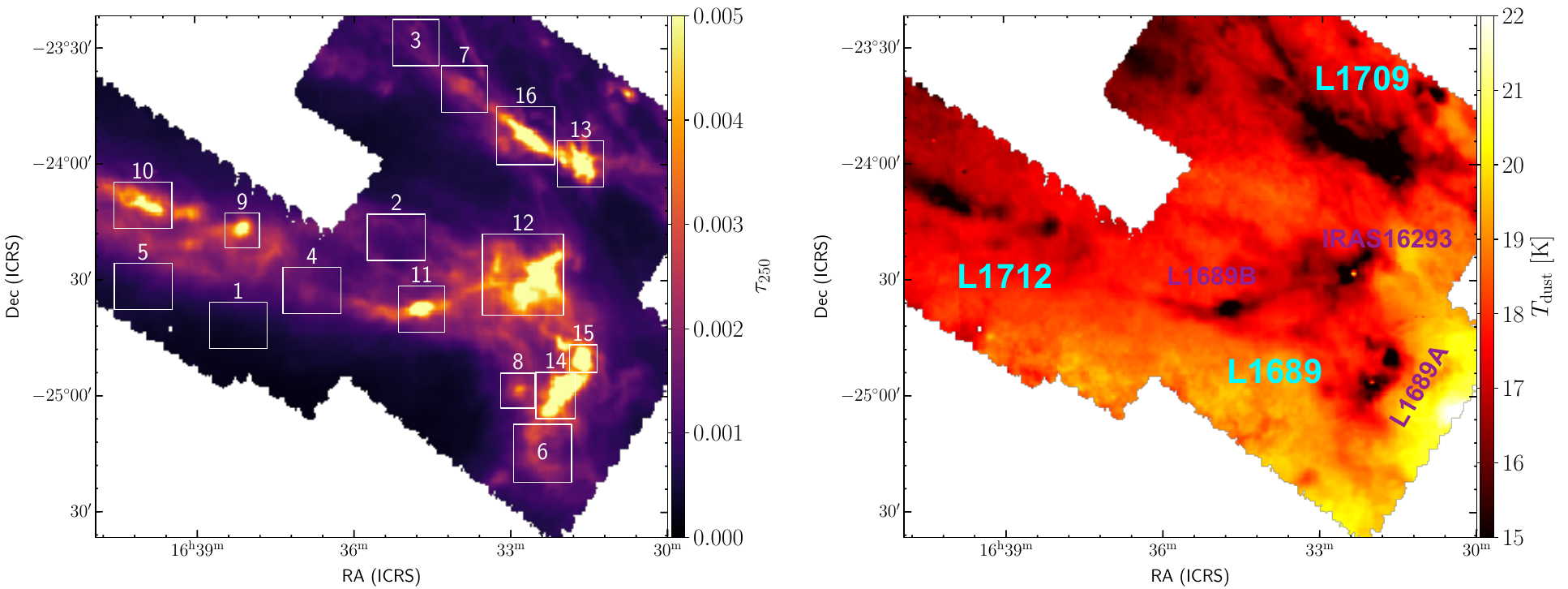}
	\caption{Maps of optical depth at $250\mum$ $\tau_{250}$ (\textit{left}) and dust temperature $T_{\rm d}$ (\textit{right}) for the Ophiuchus cloud derived from the pixel-by-pixel SED fitting. The white rectangles overlaid on the $\tau_{250}$ map indicate the sub-regions which are analyzed in Section \ref{subsec:opacity}. 
	\label{fig:oph_tau_map} }
\end{figure*}

Adopting $\lambda_0=250\mum$ ($\nu_0=1200\ghz$) as the reference wavelength, Equation \ref{equ:sed} can be rewritten as a modified blackbody: $I_\nu=B_\nu(T_{\rm d})\tau_{250}(\nu/1200\ghz)^\beta$. Here we fix $\beta$ at 2 while treating $T_{\rm d}$ and $\tau_{250}$ as free parameters. A $\chi^2$ minimization technique is then applied to fit the SED for each pixel in the calibrated \herschel\ maps. The errors in SED fitting parameters are estimated using the Monte Carlo (MC) method. All SEDs are weighted by $\sigma^{-2}$, where $\sigma$ is the sum of the squares of (i) the uncertainty of the zero correction obtained from \planck\ model, (ii) the mean Root Mean Square  (RMS) noise at a resolution of 1 arcmin ($\rm RMS_{1'}$) estimated from the \herschel\ noise maps. Calibration errors and color correction errors are accounted for in the MC analysis. Table \ref{tab:sed_cal} lists the relevant parameters used in SED fitting, including flux calibrations, color corrections, zero point ($a_\nu$) and $\rm RMS_{1'}$. A 7\% uncertainty in flux calibration for all five bands is adopted according to the PACS and SPIRE manuals (PACS Observer’s Manual, Version 2.4; SPIRE Observers’ Manual, Version 2.4).
The color corrections and uncertainties are refered to \cite{Pezzuto2012} and \cite{Sadavoy2013}. Figure \ref{fig:sed_fit_example} illustrates a representative SED and best-fit model. Figure \ref{fig:oph_tau_map} shows the optical depth ($\tau_{250}$) and dust temperature ($T_{\rm d}$) maps for the dark clouds L1689, L1709 and L1712. Generally, regions with lower temperatures correlate with regions of higher optical depths. The derived typical values of the $1\sigma$ relative errors of $T_{\rm d}$ and $\tau_{250}$ are 3.5\,\% and 15\,\%, respectively. 

\section{Results and Discussions} \label{sec:result}

\subsection{Determination of Opacity}\label{subsec:opacity}

Interstellar dust is homogeneously mixed with gas across a variety of environments, implying that the dust-to-gas ratio does not exhibit significant variation with respect to density or the sightline direction. The ratio of dust extinction or reddening to gas column density remains relatively constant across a wide range of conditions \citep{Zhu2017}. According to Equation \ref{equ:tau_kappa}, by correlating the two independent measures of $A_K$ and $\tau_{250}$ outlined in Section \ref{sec:ana}, and with the employment of an empirically calibrated ratio that connects $A_K$ to $N_{\rm H}$, the dust opacity $r\kappa_{250}$ can be determined as follows:
\begin{equation}\label{equ:rkappa}
    r\kappa_{250}=\frac{1}{\mu m_{\rm H} N_{\rm H}/A_K} \frac{\tau_{250}}{A_K},
\end{equation}
which is dependent on the empirical ratio of $N_{\rm H}/A_K$. In the typical diffuse ISM, the ratio of $N_{\rm H}$ to $A_V$ is estimated to be approximately $1.9 \times 10^{21}\,\rm cm^{-2}\,mag^{-1}$ \citep{Bohlin1978}. By applying the extinction law with $R_V=3.1$ from the models of \cite{weingartner2001} (hereafter WD01), where $A_K/A_V=0.11$, the corresponding $N_{\rm H}/A_K$ is referred to be about $1.7\times 10^{22}\,\rm cm^{-2}\,mag^{-1}$. For dense clouds, \cite{Vuong2003} reports  $N_{\rm H}/A_J \approx (5.6-7.2) \times 10^{21}\,\rm cm^{-2}\,mag^{-1}$ toward $\rho$ Ophiuchi cloud, while \cite{martin2012} gives $N_{\rm H}/E(J-K) = (11.5 \pm 0.5) \times 10^{21}\,\rm cm^{-2}\,mag^{-1}$ in the Vela cloud. Using the ratio $A_J/A_K=2.5$ \citep{Indebetouw2005}, these ratios suggest an $N_{\rm H}/A_K$ range of approximately $(1.4-1.8)\times 10^{22}\,\rm cm^{-2}\,mag^{-1}$. Given this context, $N_{\rm H}/A_K=1.7\times10^{22}\,\rm cm^{-2}\,mag^{-1}$ is adopted in this work.

\begin{figure*}
	\centering
	\includegraphics[scale=0.5]{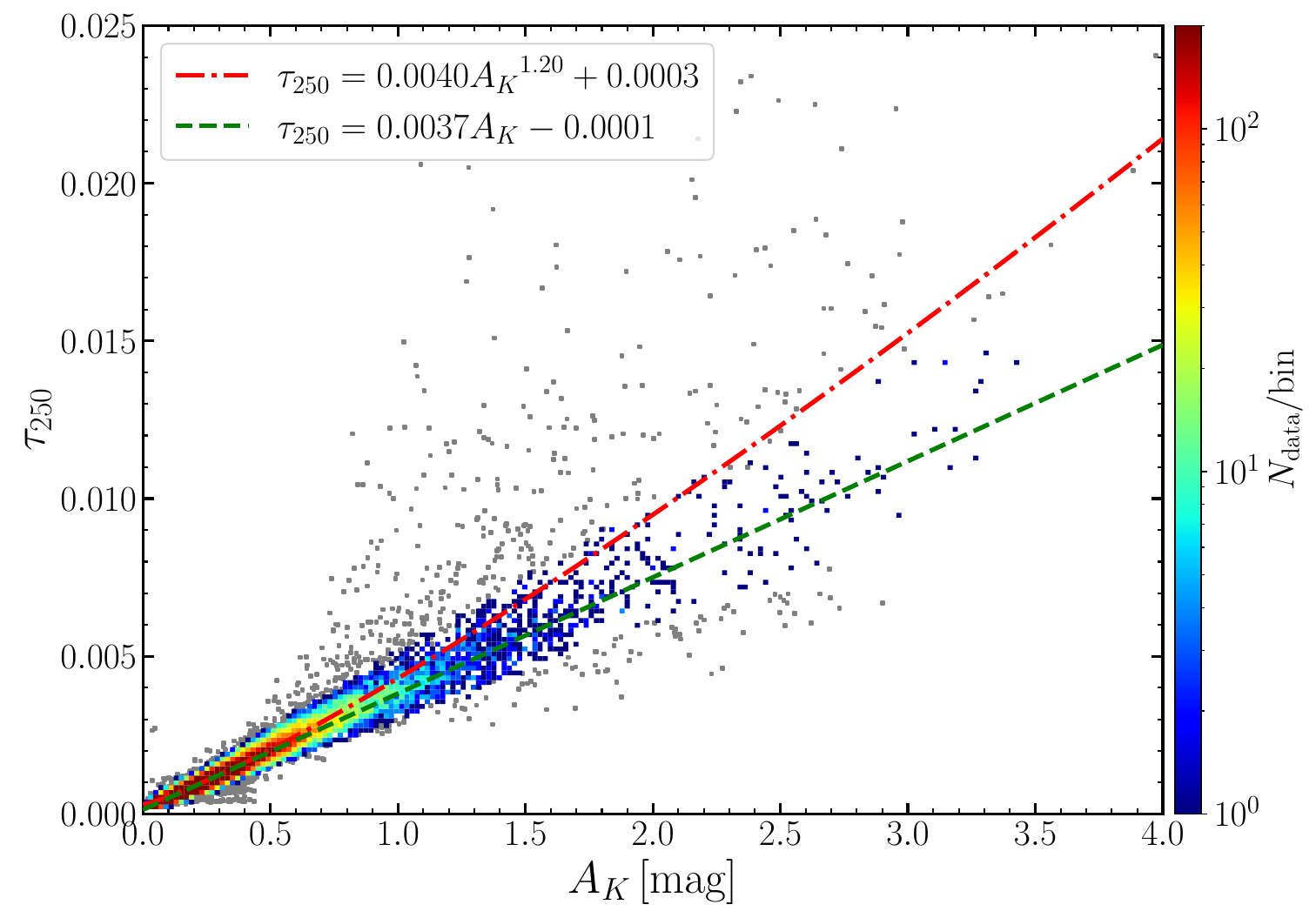}
	\caption{Correlation between FIR optical depth $\tau_{250}$ and NIR extinction $A_K$ in three dark clouds L1689, L1709 and L1712. The green dashed line presents a linear fit to the data point using the \textsc{lts\_linefit} routine. The grey points shows the outliers as determined by \textsc{lts\_linefit}, which are excluded from the linear fit. The red dot-dashed indicates the power-law correlation obtained by fitting all the data points. The best-fit parameters are displayed in the plot. 
	\label{fig:tau_ak_total} }
\end{figure*}

Figure \ref{fig:tau_ak_total} illustrates a clear correlation between $A_K$ and $\tau_{250}$ across the entire region, after excluding data points with SNRs below 3. Notably, there is a larger scatter for $A_K>0.6$ mag, yet the overall correlation between $A_K$ and $\tau_{250}$ remains apparent. This increased dispersion at higher extinctions may suggest greater variations in dust opacity along the line of sight, particularly toward the densest core regions. To estimate the overall dust opacity $r\kappa_{250}$, a linear regression is conducted on the observed correlation between $\tau_{250}$ and $A_K$. For this purpose, the Python package \textsc{lts\_linefit}\footnote{https://pypi.org/project/ltsfit/}\citep{Cappellari2013} was used, which takes into account errors in both variables, potential outliers and intrinsic scatter. As shown in Figure  \ref{fig:tau_ak_total}, the grey points represent the outliers as determined by \textsc{lts\_linefit}. The regression yielded a slope of $\tau_{250}/A_K=0.0037$, corresponding to a mean opacity value of $r\kappa_{250}=0.093\cmg$​ throughout the entire region. Considering a power-law variation of $\tau_{250}$ and $A_K$ found in the literature, we fitted a power-law function to all data points, yielding the relationship $\tau_{250}=0.004{A_K}^{1.20}+0.003$. This result demonstrates that $\tau_{250}$ increases with $A_K$ (or $N_{\rm H}$), to a power of approximately 1.2, implying that $r\kappa_{250}$ increases with $A_K$ to the power of 0.2 (see Equation \ref{equ:rkappa}). This dependency indicates potential grain growth as the cloud density increases. In comparative literature, \cite{roy2013} reported a power-law index of $1.28\pm0.01$ in Orion A, while \cite{Okamoto2017} discovered an index of 1.32$\pm$0.04 for Perseus. More recently, \cite{Hayashi2019} reported a value of 1.21$\pm$0.04 for Chamaeleon. Although the power-law index of 1.20 derived in this work is close to these findings in other molecular clouds, the power-law relationship between $\tau_{250}$ and $A_K$ is relatively weaker in comparison to a linear relationship.

It is notable that these studies as mentioned above primarily used 2MASS extinction maps that have a comparatively low resolution of approximately $3'-5'$, rendering the resolving of smaller structures challenging.  Despite our extinction mapping achieving a threefold improvement in resolution, the overall results exhibit no significant alterations. A potential cause could be that the large dispersion among data points with high extinction leads to inaccuracies in the fitting results. One might hypothesize that the resolution of finer details would uncover more pronounced density gradients, potentially leading to larger deviations in the $\tau_{250}$ and $A_K$ relationship from a simple linear approximation. However, the continuous presence of the shallow power-law, even at higher resolutions, suggests a smooth transition from diffuse to dense conditions over scales larger than those captured by our current observations. Another consideration involves beam dilution in both FIR emission and extinction maps which could average out inherent small-scale variations. Further high-resolution mapping of approximately 10$''$ would reveal localized opacity variations. Nevertheless, the consistency between our $\tau_{250}$-$A_K$ results and those from previous lower-resolution studies suggests that the inferred dust evolutionary trends remain reliable across an extensive scale range.

Considering the complexity of various environments within the Ophiuchus molecular cloud, the entire region is divided into 16 distinct sub-regions to analyze the spatial variations. The positions of these sub-regions are illustrated in Figure \ref{fig:oph_tau_map}. The typical sizes of these sub-regions are $\sim15'$, corresponding to a physical scale of $\sim$0.6 pc. In order to derive the opacity $r\kappa_{250}$ via Equation \ref{equ:rkappa}, a linear fit is performed to the $\tau_{250}$ versus $A_K$ relation in each sub-region by \textsc{lts\_linefit}. This fitting approach provides robust $\tau_{250}$-$A_K$ correlations for each sub-region. The resultant FIR opacities $r\kappa_{250}$ are listed in Table \ref{tab:opacity}, spanning a range from 0.07 to 0.12\cmg. Table \ref{tab:opacity} also presents the derived $\sigma_{250}$, along with the mean dust temperature $\langle T_{\rm d}\rangle$ and the average extinction $\langle A_K\rangle$  for each sub-region.

\begin{table*}
\centering
\caption{Basic Properties of the 16 Sub-regions and Measured Dust Opacities.}
\label{tab:opacity}
\begin{tabular}{c c c c c c}
\hline \hline
Sub-region$^a$ & $\langle T_{\rm d} \rangle$ & $\langle A_K\rangle$ & $A_K$ range & $r\kappa_{250}$ & $\sigma_{250}$\\
 & (K) & (mag) & (mag) & (\cmg) & ($\times10^{-25}\,\rm cm^2\,H^{-1}$)\\
\hline
 1 &18.40$\pm$0.22 &0.10$\pm$0.06 &0.02/0.27 & 0.075$\pm$0.002  & 1.77$\pm$0.05\\
 2 &17.80$\pm$0.24 &0.29$\pm$0.05 &0.18/0.42 & 0.078$\pm$0.005  & 1.83$\pm$0.10\\
 3 &16.76$\pm$0.27 &0.26$\pm$0.06 &0.13/0.43 & 0.082$\pm$0.004  & 1.92$\pm$0.08\\
 4 &17.90$\pm$0.20 &0.34$\pm$0.06&0.19/0.49 & 0.085$\pm$0.004  & 1.98$\pm$0.09\\
 5 &17.82$\pm$0.19 &0.19$\pm$0.10&0.03/0.49 & 0.070$\pm$0.002  & 1.63$\pm$0.04\\
 6 &18.87$\pm$0.37 & 0.39$\pm$0.11 &0.19/0.71 & 0.105$\pm$0.003  & 2.45$\pm$0.06\\
 7 &16.39$\pm$0.42 &0.38$\pm$0.16 &0.07/0.88 & 0.076$\pm$0.002  & 1.79$\pm$0.04\\
 8 &18.05$\pm$0.51 & 0.47$\pm$0.16&0.21/1.03 & 0.090$\pm$0.003  & 2.10$\pm$0.07\\
 9 & 16.74$\pm$0.59&0.66$\pm$0.32 &0.04/1.83 & 0.093$\pm$0.003  & 2.17$\pm$0.06\\
 10 &16.22$\pm$0.88 &0.63$\pm$0.27 &0.18/1.88 & 0.098$\pm$0.002  & 2.30$\pm$0.04\\
 11 &16.90$\pm$1.06 &0.64$\pm$0.46 &0.21/2.73 & 0.086$\pm$0.002  & 2.00$\pm$0.04\\
 12 &16.79$\pm$0.77 & 0.93$\pm$0.49 &0.23/3.14 & 0.095$\pm$0.001  & 2.23$\pm$0.02\\
 13 &14.67$\pm$1.34 & 0.83$\pm$0.57 &0.16/3.32 & 0.093$\pm$0.001  & 2.17$\pm$0.03\\
 14 &16.92$\pm$0.90 & 1.07$\pm$0.61&0.37/3.56 & 0.099$\pm$0.002  & 2.33$\pm$0.05\\
 15 &16.40$\pm$1.29 & 1.09$\pm$0.63&0.39/4.09 & 0.121$\pm$0.005  & 2.83$\pm$0.11\\
 16 &15.18$\pm$1.38 & 0.64$\pm$0.55&0.08/5.70 & 0.094$\pm$0.001  & 2.20$\pm$0.03\\

\hline
\end{tabular}
\tablecomments{(a) The positions of sub-regions are outlined in Figure \ref{fig:oph_tau_map}. }
\end{table*}

\subsection{Comparison with Previous Studies}

Dust opacity can be expressed in a variety of forms across diverse wavelengths. In the present work, we represent the opacity in terms of the opacity at 250 $\mu\rm m$, denoted as $r\kappa_{250}$. In Section \ref{subsec:opacity}, we derived an average opacity $r\kappa_{250} \approx 0.093\cmg$ encompassing the entire region under investigation. When analyzing the sub-regions individually, we observed $r\kappa_{250}$ values ranging from $0.07\cmg$ to $0.12\cmg$. The diffuse ISM at high latitudes typically exhibits an opacity around $\tau_{250}/N_{\rm H} \approx \rm 1\times 10^{-25}\,cm^2\,H^{-1}$ \citep{Boulanger1996}. Observations from the \planck\ mission suggest $\tau_{250}/N_{\rm H}$ values within the range of $\rm (0.6-1.6)\times 10^{-25}\,cm^2\,H^{-1}$ \citep{Planck2011A&A...536A..24P,Planck2014A&A...566A..55P,Planck2014A&A...571A..11P}, which correspond to $r\kappa_{250}$ values between 0.025 and 0.068\cmg. These findings align with the predicted $r\kappa_{250} = 0.047\cmg$ from the dust model of \cite{Draine2007}. The mean value derived in this work is approximately twice the value for the diffuse ISM. In regions 2 and 4 with lowest extinction of $A_K<0.2$\,mag, we determined $r\kappa_{250} \approx 0.07\cmg$, larger than that in the diffuse ISM. The NIR extinction map in this work is insensitive to the diffuse region with $A_K<$ 0.1 mag. As indicated in Figure \ref{fig:hist_ext}, the sensitivity of the derived extinction map, denoted as $\sigma_{A_{K}}$, ranges from 0.02 to 0.1 mag.

Instead of extinction mapping, \cite{suutarinen2013} employed the use of smoothed color excess $E(H-K)$ derived from NTT/SOFI observations targeting the prestellar core CrA C. They found an opacity of $r\kappa_{250} = 0.08 \pm 0.01\cmg$. This value does not demonstrate a significant enhancement when compared to the diffuse ISM, to some extent, it is less than most of the results obtained in this work.

\cite{lewis2022} conducted a comparison between the 2MASS extinction map with \planck\ emission at a resolution of 5 arcmin for the Ophiuchus cloud. Their derived opacity $r\kappa_{250}\approx0.07\cmg$ is lower than our average value and even the minimum value among the sub-regions. It is worth noting that the adopted extinction law could influence the resultant opacity. Flatter curves tend to yield higher extinctions from color excess measurements. For instance, varying between the $R_V$=3.1 and $R_V$=5.5 extinction laws from WD01 introduces uncertainties up to $\sim$20\% in $A_K$ and consequently in $r\kappa_{250}$. In addition, a lower spectral index $\beta$ assumed in SED fitting results in lower $\tau_{250}$, which subsequently lead to decreasing $r\kappa_{250}$ by approximately 25\% for $\beta$=1.8 compared to $\beta$=2.2 \citep{suutarinen2013}. However, the primary factor contributing to the lower opacity in \cite{lewis2022} is the limited spatial resolution. Their resolution of 5 arcmin ($\sim$0.2\,pc) biases their opacity estimate towards nearby translucent ISM. As a result, the $r\kappa_{250}$ value that we derive is more reflective of values in dense environments where dust grain evolution is expected to be significant. Our results emphasize the necessity of high resolution in both extinction and emission maps to accurately characterize opacity variations related to local density enhancements.

\subsection{Opacity as a Tracer of Dust Evolution}

Dust grains at various stages of evolution display distinct FIR opacity characteristics. Specifically, an increase in FIR opacity often indicates grain growth, as larger dust grains contribute more significantly to the opacity at longer wavelengths \citep{Kohler2012,Kohler2015,ossenkopf1994,ormel2011}. The grain growth is also often observed in denser regions of the ISM and in protoplanetary disks, where dust grains can accumulate and coagulate \citep{Testi2014}. According to the models proposed by \cite{ossenkopf1994}, ISM-like dust without ice mantles and coagulation has an opacity $r\kappa_{250} \sim 0.06$ \cmg, which increases to 0.13 \cmg\ when dust grains evolve over a coagulation time of $10^5$ years at a gas density of $10^5$ $\rm cm^{-3}$. Under similar conditions, the updated models by \cite{ormel2011} predict opacities reaching 0.14-0.17 \cmg\ for fully ice-coated aggregates \texttt{(ic-sil, ic-gra)} or spatially mixed aggregates \texttt{(is-sil+gra)}. Moreover, with greater coagulation time and density, both models can yield even higher opacities (up to 0.2 \cmg\ or more). Therefore, we deduce that the opacities observed in the Ophiuchus cloud can be consistent with these models under specific conditions.

Numerous studies have previously identified the trend of increasing FIR dust opacity with higher environment density by correlating the FIR optical depth with extinction. For instance, \cite{kramer2003} observed increasing opacity with decreasing temperature in IC5146. \cite{Ysard2013} found a factor of two higher opacity in the center compared to the outskirts of the L1506 filament in Taurus. \cite{Stepnik2003} utilizing PRONAOS observations of Taurus measured a 3.5 times increase in $\tau_{250}/N_{\rm H}$. \cite{Schnee2008} also reported higher emissivities for colder grains. Dust coagulation can account for the opacity enhancements and temperature drops in dense regions. Radiative transfer models indicate this opacity rise cannot be explained by changes in the radiation field alone, but instead reflects variations in intrinsic dust properties \citep{juvela2015,Ysard2013,suutarinen2013}. While many observations indicate grain growth in dense regions, \cite{suutarinen2013} did not find clear opacity enhancement in a prestellar core in CrA. This suggests that opacity alone does not definitively trace evolution. Therefore, complimentary constraints from dust scattering properties are also needed to quantify grain growth \citep{Juvela2020}.

\begin{figure*}
	\centering
	\includegraphics[scale=0.65]{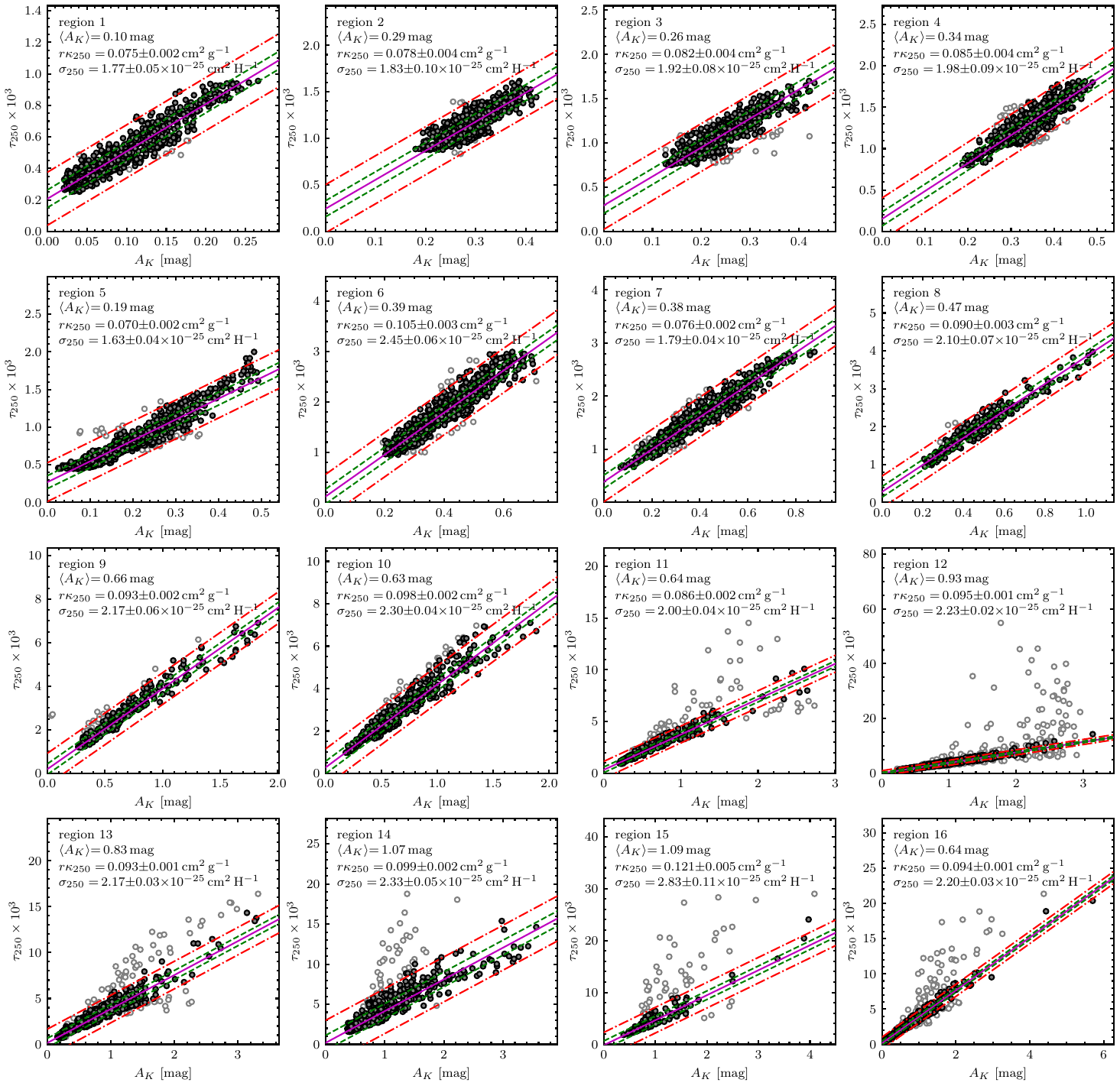}
	\caption{Pixel-by-pixel correlation of the optical depth $\tau_{250}$ and extinction $A_K$ for all sub-regions defined in Figure \ref{fig:oph_tau_map}. The blue solid, green dashed, and red dot-dashed lines indicate the best-fit, 1σ, and 3σ lines, respectively, obtained using the \textsc{lts\_linefit} fitting procedure. The derived parameters of average extinction, $r\kappa_{250}$ and $\sigma_{250}$ from the best-fit line are shown in the panels.  Open circles are the outliers excluded from the fitting automatically. Note that the axis scales differ between panels to accommodate the variable dynamic ranges of $A_K$ in each sub-region.
	\label{fig:rkappa_subregion} }
\end{figure*}

Figure \ref{fig:rkappa_subregion} presents the correlations between $\tau_{250}$ and $A_K$ across all 16 sub-regions as defined in Figure \ref{fig:oph_tau_map}. The majority of these sub-regions exhibit a tight linear relationship of $\tau_{250}$ and $A_K$.  However, in some regions with higher dynamic range of $A_K$ (e.g. regions 11-16), the  \textsc{lts\_linefit} linear fitting technique has detected a significant number of outliers. These outliers primarily deviate above the best-fit linear relationship at higher $A_K$ values, which may suggest a potential increase in dust opacity that could be indicative of grain growth. It is also plausible that these outliers are attributed to the underestimation of extinction toward the dense cores, potentially due to a limited number of background stars. Moreover, these dense cores are too small to be resolved with the 1 arcmin beam in our extinction mapping, leading to larger dispersion when $A_K>1.5$\,mag. The localized opacity increase toward dense cores merits further investigation through high angular resolution observations that can resolve the small-scale structure. To distinguish potential opacity variations from mapping artifacts, it is crucial to match the resolution between the FIR emission and NIR extinction measurements.

\begin{figure*}
	\centering
	\includegraphics[scale=0.65]{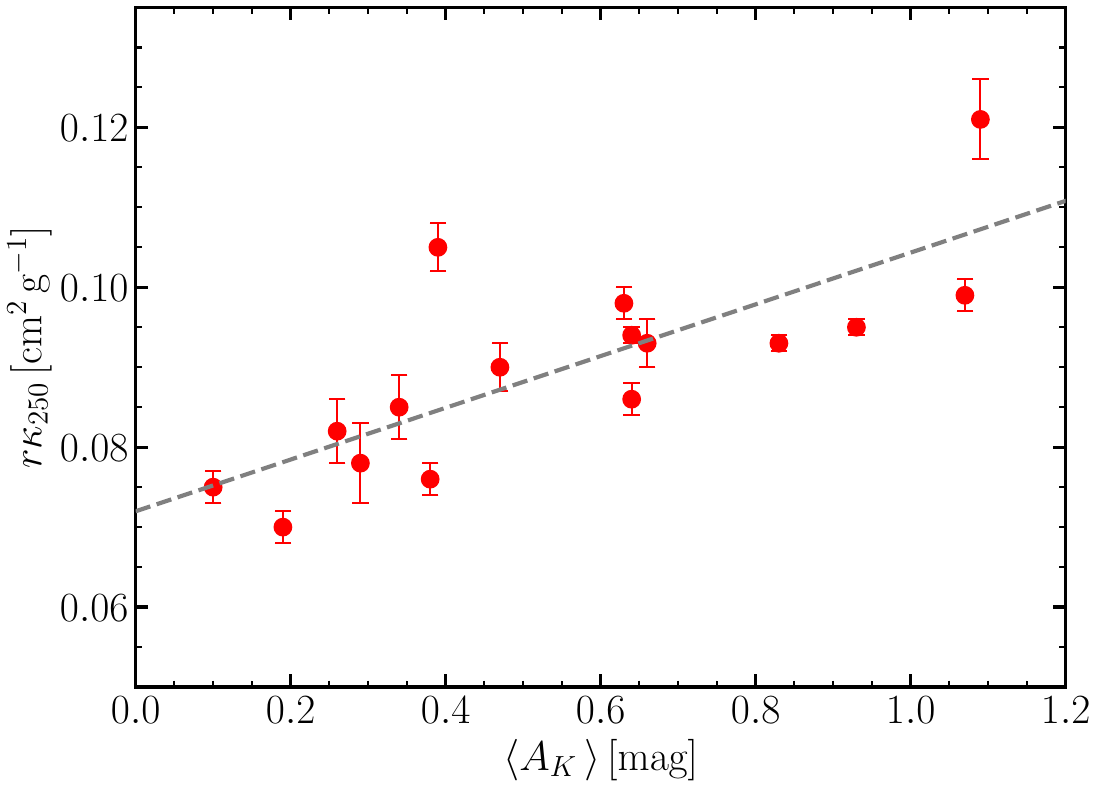}
	\caption{The opacity $r\kappa_{250}$ as a functin of average extinction $\langle A_K\rangle$ for the 16 sub-regions listed in Table \ref{tab:opacity}. The dashed line represents a linear fit to the data, offering direct evidence of grain growth.
	\label{fig:rkappa_ak} }
\end{figure*}

As shown in Figure \ref{fig:rkappa_ak}, the overall variation in opacity across the region is within a factor of two. Nevertheless, there is a discernible general trend of increasing opacity with dust column density, particularly towards cloud cores where grain growth appears enhanced. Despite the complexities introduced by including both lower density and dense star-forming cores, the opacities remain relatively uniform, only exhibiting a factor of two variation across the entire region. The lack of strong opacity enhancements even toward dense cores implies rapid grain growth has not yet dramatically altered the dust properties in most areas. The handful of sub-regions with $r\kappa_{250}$ exceeding 0.1\cmg\ (see Figure \ref{fig:rkappa_subregion} and \ref{fig:rkappa_ak}) likely reflect grain coagulation in the cold dense cores. But overall, the narrow distribution suggests the dust populations in the Ophiuchus dark clouds are likely at similar evolutionary stages, with no clear signs of rapid grain growth.


%


\section{Summary} \label{sec:summary}

In this work, we use multi-wavelength infrared observations from UKIDSS (1-3\mum), \spitzer\ (3-8\mum), and \herschel\ (100-500\mum) to expore dust opacity in three dark clouds (L1689, L1709, and L1712) within the Ophiuchus cloud complex. Our primary conclusions are as follows:

\begin{enumerate}
	\item{By integrating the NICE and NICER extinction measurement methods and leveraging NIR and MIR photometry, we generate extinction maps for the three dark clouds at a resolution of 1 arcmin (equivalent to 0.04 pc) and depth up to $A_K \sim 3$ mag. This resolution and depth exceed previous of prior 2MASS-based extinction maps by more than threefold, thereby enabling the resolution of distinct dense cores and filaments within the clouds.}
	
	\item{Comparing the $A_K$ extinction with \herschel\ FIR emission allows us to determine the overall dust opacity of the Ophiuchus cloud to be $r\kappa_{250} \approx 0.09\cmg$ on average, which is approximately 2-3 times higher than that of the diffuse ISM. This is consistent within errors with previous measurements based on 2MASS extinction map \citep[e.g.][]{martin2012,roy2013}. It can be expected that observations by the James Webb Space Telescope (JWST), combined with submillimeter observations will extend this method and determine the dust opacity of interstellar clouds with unprecedented precision.}

	\item{It is found that the dust opacity $r\kappa_{250}$ increases with density ($A_K$ or $N_{\rm H}$) as a power law with an index of $\sim$0.2, suggesting grain growth. An analysis of sub-regions indicates that $r\kappa_{250}$ also exhibits spatial variations, generally increasing with higher environmental density. Although there are indications of grain growth, the evidence does not suggest rapid grain growth has taken place in the dark clouds examined in this study. }
	
\end{enumerate}

\begin{acknowledgments}
We would like to thank the anonymous referee for the very helpful comments and suggestions. We thank Dr. Yuping Tang for helpful discussions. This work is supported by the National Key R\&D program of China (2022YFA1603102), the National Natural Science Foundation of China (12133002). H.Z. is funded by the China Postdoctoral Science Foundation (No. 2022M723373) and the Jiangsu Funding Program for Excellent Postdoctoral Talent. X.C. thanks to Guangdong Province Universities and Colleges Pearl River Scholar Funded Scheme (2019).

\end{acknowledgments}

%

\vspace{5mm}
\facilities{UKIRT, \spitzer(IRAC), \herschel(PACS and SPIRE)}


\software{Astropy \citep{astropy2018},
        APLpy \citep{aplpy2012}, LtsFit \citep{Cappellari2014}, TOPCAT \citep{Taylor2005}.
          }




\bibliography{opacity-Oph}{}
\bibliographystyle{aasjournal}

\end{CJK*}
\end{document}